\documentclass[twocolumn]{aa}
\usepackage{amsmath}
\usepackage{amssymb}
\usepackage{epsfig}
\usepackage{txfonts}
\usepackage{natbib}
\usepackage{graphicx}
\usepackage{float}
\usepackage{here}
\usepackage{subfig} 
\usepackage[figuresright]{rotating}

\newcommand{\arcs}		{$^{\prime\prime}$}
\newcommand{\arcm}		{$^{\prime}$}

\newcommand{\mum}		{$\mu$m}
\newcommand{\msun}		{M$_{\odot}$}

\newcommand{\jl}		{$J$}
\newcommand{\hl}		{$H$}
\newcommand{\kl}		{$K_{\rm{S}}$}
\newcommand{\hk}		{$(H-K_{\rm{S}})$}

\newcommand{\gt}		{$>$}
\newcommand{\lt}		{$<$}

\begin{document}
   \title{Searching for dark clouds in the outer galactic plane}

   \subtitle{I -- A statistical approach for identifying extended red(dened) regions in 2MASS}

   \author{Wilfred W.F. Frieswijk\inst{1}
          \and
          Russell F. Shipman\inst{2,1}
          }

   \offprints{W.F.F. Frieswijk}

   \institute{Kapteyn Astronomical Institute, University of Groningen, PO Box 800, Landleven 12, 9700\ \ AV Groningen, The Netherlands\\
              \email{frieswyk@astro.rug.nl}
         \and
            SRON, National Institute for Space Research, PO Box 800, Landleven 12, 9700\ \ AV Groningen, The Netherlands\\
             }

   \date{Received - -, -; accepted - -, -}

   \abstract{Most of what is known about clustered star formation to date comes 
from well studied star forming regions located relatively nearby, such as 
Rho-Ophiuchus, Serpens and Perseus.
However, the recent discovery of infrared dark clouds may give new insights in our 
understanding of this dominant mode of star formation in the Galaxy. 
Though the exact role of infrared dark clouds in the formation process is still somewhat 
unclear, they seem to provide useful laboratories to study the very early stages of 
clustered star formation.
Infrared dark clouds have been identified predominantly toward the bright inner parts of 
the galactic plane. The low background emission makes it more difficult
to identify similar objects in mid-infrared absorption in the outer parts.
This is unfortunate, because the outer Galaxy represents the only nearby region 
where we can study effects of different (external) conditions on the star formation 
process.}
{The aim of this paper is to identify extended red regions in the outer galactic plane
based on reddening of stars in the near-infrared. We argue that
these regions appear reddened mainly due to extinction caused by molecular clouds
and young stellar objects.
The work presented here is used as a basis for identifying star forming regions and 
in particular the very early stages. An accompanying paper describes the 
cross-identification of the identified regions with existing data, uncovering more on 
the nature of the reddening.}
{We use the Mann-Whitney U-test, in combination with a friends-of-friends algorithm,
to identify extended reddened regions in the 2MASS all-sky $JHK$ survey. We process 
the data on a regular grid using two different resolutions, 60\arcsec and 90\arcsec.
The two resolutions have been chosen because the stellar surface density varies between
the crowded spiral arm regions and the sparsely populated galactic anti-center region.}
{We identify 1320 extended red regions at the higher resolution and 1589 at the lower 
resolution run. The linear extent of the identified regions ranges from a few arc-minutes
to about a degree.}
{The majority of extended red regions are associated with major molecular cloud 
complexes, supporting our hypothesis that the reddening is mostly due to foreground 
clouds and embedded objects. The reliability of the identified regions is $>$99.9\%. 
Because we choose to identify object with a high reliability we can not 
quantify the completeness of the list of regions.}
 
     \keywords{ Method: statistical -- Astronomical data bases: Catalogs -- Stars: formation -- 
     ISM: clouds -- dust, extinction }

\authorrunning{W.W.F. Frieswijk and R.F. Shipman}
\titlerunning{Extended red regions in the outer galactic plane}
   \maketitle
\section{Introduction}
\label{intro}
Dark clouds represent the earliest stages of star formation in the Galaxy.
Nearby dark clouds have been studied in great detail and provide a wealth
of information on the low-mass isolated star formation process 
\citep[e.g.,][]{Shu1987}. However, most stars do not form
in isolation but in groups or clusters \citep{Lada2003}, and this
clustered mode of star formation is not well understood. Some key questions
of current star formation research are the origin of the stellar mass 
distribution, the relation between high-mass star formation and stellar 
clusters and the effect of environment on the star formation process. 
See \citet{Zinnecker2007} for a recent review of this topic.

In the mid 1990s, infrared surveys with the Infrared Space Observatory (ISO)
and the Midcourse Space Experiment (MSX) have revealed a class of dark clouds, 
referred to as Infrared Dark Clouds \citep[IRDCs][]{Perault1996,Egan1998}.
These clouds are observed in silhouette against the bright infrared background 
emission of the galactic plane.
IRDCs are typically cold ($<$25\,K) and dense ($\sim$10$^5$\,cm$^{-3}$) with masses ranging
from \citep[$\sim$100--10$^5$\,\msun;][]{Egan1998,Carey1998,Carey2000} and they are 
located at larger distances ($>$1\,kpc) compared to well-studied low-mass star forming 
regions.
A recently published catalog of dark clouds extracted from the GLIMPSE Survey by 
\citet{Peretto2009} shows 
that most IRDCs have column densities below N(H$_2$)$\approx$5$\times$10$^{22}$\,cm$^{-2}$, 
with some extreme objects reaching N(H$_2$)$>$10$^{23}$\,cm$^{-2}$.
This new class of dark clouds are believed to represent the very 
early stages of clustered star formation, hence commonly referred to as cluster-forming 
clumps \cite[e.g.,][]{Rathborne2006}.

While most of the star-forming activity in the Galaxy takes place in the inner 
spiral arms and the molecular ring, significant activity also occurs beyond the solar 
circle (R$_g$ $>$ 8.5\,kpc). Prominent examples include well studied nearby ($<$500\,pc) 
regions such as the Orion and Taurus-Perseus-Auriga complexes as well as more distant 
($>$2\,kpc) active star forming regions such as \object{W3} and \object{NGC 7538}. 
However, the outer Galaxy is generally somewhat neglected in studies of star formation 
compared with the inner Galaxy, which is unfortunate because it is a useful laboratory 
to study the effect of different external conditions on the process that initiates stellar 
birth. In particular, the metallicity, pressure, and radiation field in the outer Galaxy are 
considerably different from the inner Galaxy \citep{Brand1995,Rudolph2006}, which makes 
it by far the most nearby region to study the effect of such variations on star 
formation.

The aim of this work is to locate star-forming regions in the outer galactic
plane. To find concentrations of molecular gas in the outer galactic plane,
where the mid-infrared background is insufficient to measure absorption, we use 
the extinction to background stars. The traditional method to
do this is via star counts \citep[e.g.,][]{Wolf1923}, which is limited by extinction 
to nearby clouds, especially when using optical data. This problem is avoided by using
near-infrared data, but pick-up of emission from embedded objects makes this
method unreliable.

An alternative to star counts is to use the near-infrared colour excess 
\citep[NICE,][]{Lada1994} which is widely used to map the extinction toward dark clouds 
\citep[e.g.,][]{Cambresy2002,Alves1998,Lada1999}. However, this method relies on knowledge 
of the distance to the object, so that a correction for foreground stars can be made.
The presence of foreground stars can reduce the measured color excess, and hence may
influence an identification based on this, significantly \citep{Lombardi2005}.
Furthermore, identifying objects in an automated, uniform way toward large regions 
in the sky is difficult using NICE because the extinction law is not uniform in 
every direction. Also, an average reference color of background stars is required, 
which is usually adopted from a nearby, well selected, extinction-free region.\\
\ \\
The identification of dark clouds by the extinction they impose upon
background stars requires a method that:
\begin{list}{}{\topsep=5pt\itemsep=0pt\leftmargin=25pt\rightmargin=15pt\itemindent=0pt}
	\item[1)]determines the extinction from the colors of stars,
	\item[2)]can be applied to small numbers of stars (for high resolution),
	\item[3)]can be treated uniformly in all directions,
	\item[4)]puts a confidence level to measurements,
	\item[5)]is independent of foreground stars.
\end{list}
Because of the above mentioned limitations when using a near-infrared color excess method, 
we have chosen to search for candidate dark cloud regions in a statistical manner rather 
than based on the absolute value of the extinction.

In two papers, we describe the identification process of candidate regions in the outer 
galactic plane with the goal to build up a sample of objects that can be used for future 
studies of (clustered) star formation throughout the Galaxy.\\
\ \\
In this first paper we describe the Mann-Whitney U-test. This distribution-free 
statistical test can be applied uniformly to large data-sets in order to identify 
parts of the data with deviating properties. The purpose of this paper is to identify
extended red regions in the sky using data from the Two Micron All Sky Survey (2MASS).
In an accompanying paper (Frieswijk et al. 2010, hereafter Paper II) we perform 
a cross-correlation between the red regions presented 
here and existing data (optical dark clouds, MSX, IRAS and CO where available). 
The purpose of the second paper is to provide an indication of the actual nature 
of the reddening toward the identified regions

In Section \ref{data} we summarize the 2MASS data and discuss why we use the \hk\ colors 
for the identification.
In Section 3 we explain the Mann-Whitney U-test and describe the procedure that we 
follow in order to determine the statistics. 
In Section 4 we describe the friends-of-friends approach that is used to extract extended 
red regions from the reliability images produced by the U-test. 
In Section 5 we present the results. A catalog of the extended red regions is available 
online\footnote{\\
http://www.astro.rug.nl/$\sim$ismgroup/OuterGalaxy/} and at the CDS. The reliability 
images of the entire outer galactic plane can be downloaded in {\sc FITS} format 
from the same website.
The final section summarizes our main conclusions.
\section{Data: the Two Micron All Sky Survey}
\label{data}
The Two Micron All Sky Survey (2MASS) scanned the entire sky uniformly in three 
near-IR bands: \jl\ (1.25\,\mum), \hl\ (1.65\,\mum) and \kl\ (2.17\,\mum).
The 2MASS Point Source Catalog \citep[2MASS PSC,][]{Skrutskie2006} 
consists of accurate positions (astrometric accuracy RMS $<$200\,mas) and brightness 
information for over 400 million stars.
The point source catalog is \gt\,99\,$\%$ complete up to \jl\,\lt15.8, \hl\lt15.1 
and \kl\lt14.3 magnitudes. The photometrical signal-to-noise ratio is \gt\,10
(i.e. $\sigma_{J,H,K_{\rm{s}}}\lesssim0.1$\,mag) for sources brighter than the above 
stated completeness limits. The so-called faint extension of the point source 
catalog includes sources that reach 0.5 to 1.0 magnitude beyond the above limits.
The completeness, reliability and uniformity of the faint extension sources are not as 
good as the high-reliability PSC and photometric errors increase up to 
$\sigma_{J,H,K_{\rm{s}}}\sim0.4$\,magnitude.
We make use of all available sources in 2MASS, including the faint extension.
This increases the total number of objects by almost a factor 2 compared to star count 
studies, where the faint extension cannot be used because the data are incomplete.

The \hk\ color is an appropriate choice to perform the identification 
of red regions because the intrinsic \hk\ color of stars spans only a narrow range. 
For spectral types A0V to M6III the range is about 0.00 to 0.30\,mag 
\citep{Bessel1988, Wainscoat1992}. In the Solar neighbourhood the average \hk\ color 
is about 0.13\,mag with a standard deviation of 0.1\,mag. This assumes no reddening 
due to extinction and takes into account the sensitivity limit of 2MASS 
(14.3\,mag for $K_{\rm{s}}$-band). By including the faint extension, the average 
\hk\ color increases slightly (0.15\,mag) due to faint, intrinsically 
redder M-dwarfs. If a region appears significantly redder, this suggests that either 
foreground extinction is present or the intrinsic color distribution of stars in that 
direction is different. The latter can be important if for example a group of 
young stellar objects, which are intrinsically redder, is located in the observed 
field. Though such regions will be identified being intrinsically red rather than
reddened, they still represent star forming regions. A discrimination between
the two, which is not obvious from the work in this paper, can be done by
looking at other available data and will be discussed in Paper II.

We analyse the 2MASS data in the outer Galaxy from $l$=90\degr\ to 270\degr\ 
and $b$=$-$3.5\degr\ to +3.5\degr. In order to process the data on a desktop computer,
we divided the area up in regions of $2\degr \times 2\degr$, with a sampling separation 
between different fields of one degree. 
\section{Method}
\label{method}
\subsection{The Mann-Whitney U-test}
The goal of this project is to identify, based on near-IR \hk\ colors, those regions which 
are redder compared to the background. One potential approach is to calculate an 
“average” color of a few stars and test whether that average is different than a 
background sample of stars.  
This approach has some significant drawbacks: how many sample stars should be used and 
what is the stellar color of the “background” given that this value changes over the 
Galactic Plane? Is the average the correct statistic to calculate? Why not use the median 
or some other statistic?  
Furthermore, the distribution of the colors of stars is not a standard parametric 
distribution. It is definitely not a Normal distribution which is needed if we want to 
use the average colors of stars and calculate a confidence based on a standard deviation.
What is needed is a statistic, which does not depend on the distribution of stellar colors 
and which works when comparing small samples against a background of many stars in a 
robust manner.

The Mann-Whitney U-test (also referred to as the Wilcoxon rank-sum test, hereafter 
referred to as U-test) is a well-known non-parametric significance test first proposed by 
\citet{Wilcoxon1945} and extended to arbitrary sample sizes by \citet{Mann1947}. 
The statistic is based on the rank of the observation and not the observation value itself.
This makes it a non-parametric statistic. The test is used to assess whether two samples 
of observations, say sample $A$ and $B$, come from the same distribution.
Samples $A$ and $B$ are required to be independent and the observations should be ordinal or 
continuous measurements.
The U-test is hypothesis testing, where the hypothesis is that the two samples come 
from the same distribution. This is called the Null hypothesis. The statistic which is 
calculated is the sum of the joint ranking ($U$) of all the observations, in our case the 
\hk\ colors.

\noindent \emph{An example}\\
The distribution of the sum of ranks of $A$ and $B$, is readily calculated for samples 
drawn from the same parent distribution. Take the case of 3 observations against 3 other 
observations. If the observations are truly randomly drawn from the same parent 
distribution, then one would expect a lower chance of pulling the three highest values 
for sample $A$ (ranks {4,5,6}) and $B$ therefore resulting in the three lowest ranks 
{1,2,3}. One would rather expect a more “random” mix, e.g., {1,4,6} and {2,3,5}. 
In the example above the sum of the ranks in the first case, $U_A$=15, $U_B$=6 is less 
likely than the second sum of ranks, $U_A$=11, $U_B$=10. In this way the distribution 
of rank sums is built up. 

If the rank sum statistic is $U_A$=15, the probability of this occurring randomly 
is small (5\%), which means the Null hypothesis can be rejected on a statistical basis to 
a particular level of confidence (95\%), in favor of the alternative or test hypothesis 
that the $A$ samples are “larger” than the $B$ samples. The alternative hypothesis must be 
created in such a way that it is the only alternative to the Null hypothesis. 
Rejecting the Null hypothesis means that the alternative hypothesis must be accepted at 
that level of confidence.

\noindent The U-test can be applied to large scale surveys in order to identify 
regions where source properties deviate significantly with respect to a reference field.
We use the U-test statistic to test whether a set of colors of stars is different than 
the set of colors of stars from another sample. We test against the Null hypothesis that 
the set of colors is actually the same. When the test rejects the Null hypothesis 
(to a certain confidence level), we must accept the test hypothesis that the colors are 
actually different.
As can be seen from the example above, there is a fundamental difference if $U_A$ is 
“greater” or “less” than $U_B$.  For near-IR colors this means we could test for stars 
redder or bluer than the background stars.

For this work, we want to compare as few stars as possible against a background 
of tens of thousands. The benefit of using a non-parametric test is that 
whatever size the two samples have, the distribution of the U-test statistic is known
from pre-calculated tables \citep[e.g.,][]{Wall}, unlike for a parametric test. 
However, tables 
for finding the probabilities of $U$ usually do not contain values for samples with size 
in excess of 20.
For large sample sizes though, the distribution will approach a Normal distribution 
thanks to the Central Limit Theorem. The mean and standard
deviation of the variables is simply given by the mean and standard deviation of the 
Normal distribution. \citet{Moses1964} stipulates that the smallest number of samples is 
5, to still be able to use the Normal approximation for the U-test. For a parametric
test such as the average, a Normal approximation is only valid for higher sample sizes. 
This is evaluated in Section 3.5.
\subsection{Applying the U-test procedure}
We perform our calculations on a regular grid. For each grid-cell the distribution
of properties, i.e., star colors, is referred to as sample $A$. A reference 
field represents sample $B$ and is determined locally. The choice of the reference field 
is explained in Section \ref{reffield}. The size of the grid-cells can be chosen at will, 
but a minimum number of sources (5; see Sec. \ref{compare}) is required to 
retrieve a reliability based on the Normal approximation. The choice of 
the resolution is explained in more detail in Section \ref{resolution}.

\begin{figure}[!ht]
\centering
\includegraphics[width=1\columnwidth,clip]{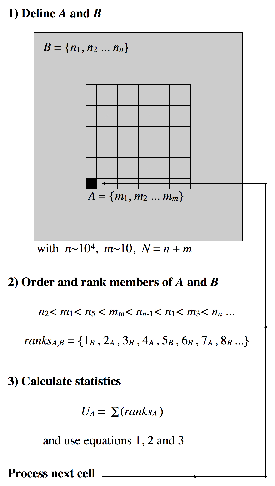}
\caption{Schematic representation of the U-test procudure.}
  \label{fig_schem}
\end{figure}

The following steps are involved in the calculation of the probability $P$ for any given 
grid-cell, where $P$ is the probability that a cell contains a redder color distribution
compared to the reference field. The whole procedure is schematically given also in 
Figure \ref{fig_schem}.
\begin{list}{}{\topsep=5pt\itemsep=0pt\leftmargin=25pt\rightmargin=15pt\itemindent=0pt}
\item[$1$)]Let $A$ be the distribution of colors in the cell (with $m$ members) and 
$B$ the local reference distribution (with $n$ members, where $n$ is very large in this 
study, i.e., $\sim$10$^{4-5}$). The Null hypothesis ($H_0$) is that $A$ and $B$ have the same parent population, 
or in perspective of our work, that the parent population of $A$ is given by $B$. Note that 
when $m$ is less than 5, the probability $P$ is set to 50\% and the next cell is 
processed.
\item[$2$)]Rank the colors in ascending order for the combined members of $A$ and $B$. 
Preserve the $A$ or $B$ identity for each member.
\item[$3$)]Sum the number of $A$-rankings to get the statistical value $U_A$. 
\end{list}
Generally, numerically equivalent (also known as tied) observations are assigned with the 
average of their respective ranks. However, we avoid tied observations by adding very small 
random numbers to the color values ($<$10$^{-4}$). This has no impact on the results, 
but makes the computations easier.
For large samples, the sampling distribution tends to a Normal distribution 
(see also Sec. \ref{compare}) with a mean $\mu_A$ and variance $\sigma_A^2$ given by
\begin{equation}
\mu_A = m(N+1)/2 \ \ \ {\rm and} \ \ \sigma_A^2 = mn(N+1)/12 ,
\end{equation}
respectively, where $N=m+n$. The significance, $z$, can then be assessed from the 
Normal distribution by calculating
\begin{equation}
z = \frac{U_A-\mu_A\pm0.5}{\sigma_A} .
\end{equation}
The term $\pm$0.5, in statistics often referred to as the continuity correction 
factor, is a correction required to maintain continuity when a discrete distribution, such as 
the U-test statistic, is approximated by a continuous distribution, such as the Normal 
distribution.
Because the Normal distribution consists of all real numbers, but the U-test values
are discreet (integer values), a Normal approximation should identify the
discreet event '$n$' with the normal interval '($n-0.5,n+0.5$)', where $n$ is any integer 
value.
In this situation we use $-$0.5 (instead of +0.5) because we consider only 
objects in the upper tail, i.e., where the distribution is redder (instead of bluer) than 
the reference distribution. The probability $P$ is calculated by evaluating the
integral from -$inf$ to $z$ of the Gaussian probability density function $\Phi(z)$, where 
$\Phi(z)$ is given by 
\begin{equation}
\Phi(z) = \frac{1}{\sqrt{2\pi}} e^{-z^2/2} .
\end{equation}

\subsection{Reference distribution}
\label{reffield}
A single reference distribution 
is almost certainly not representative 
for every field along the galactic plane. First, because the stellar population, and thus the color 
distribution, may vary with galactic coordinates. Second, because foreground extinction 
will be present everywhere. The reddening caused by the extinction depends on the amount 
of foreground material and on the dust properties toward different lines of sight.
Although the overall extinction features are interesting, the target for this work is to 
find extreme color changes relative to the local environment. Therefore we make no attempt 
to extract large scale extinction features along the galactic plane.

Instead of using a single reference distribution everywhere, we define the distribution of 
colors in every $2\degr \times 2\degr$ field as a reference for the grid cells in the inner 
square degree of the respective field. This means that we select a local reference 
distribution as best representation for the colors in each field, thus avoiding the effect 
of color variations on galactic scale.
Note that the colors of stars in the cells that are being compared to the local reference 
are omitted from the reference distribution itself.
\subsection{Resolution limitation}
\label{resolution}
The limiting factor for the spatial resolution that can be achieved in performing the
U-test is determined by the minimum of 5 sources required for the Normal approximation. 
As a requirement for the grid-cell size (resolution) we state that in every field at 
least 75\% of the cells 
should contain 5 or more sources. Because the stellar surface density decreases toward 
the galactic anti-centre ($l$=180\degr, $b$=0\degr), we process the data at two different resolutions. 
The highest possible resolution, mostly applicable towards the spiral arm regions at 
$l$$\sim$90\degr –-140\degr\ and $l$$\sim$240\degr --270\degr, is 60\arcsec.
For the region between $l$$\sim$140\degr --240\degr\ we require a grid with cells of 
90\arcsec. Note that we process the entire outer galactic plane at both resolutions 
and all data are made available.
The average number of stars is about 8 per cell for the high resolution grid toward the 
spiral arm regions as well as for the low-resolution grid toward the anti-center region.

\subsection{Minimum requirements: a Monte Carlo approach}
\label{compare}
In the U-test we assume that even for small samples drawn from a large set 
of sources, the distribution of the significance values ($z$) tends to Normal. 
With a Monte Carlo simulation we show that the significance distribution of randomly 
drawn samples from a large distribution is indeed close to a Gaussian probability density 
distribution. We specifically simulate samples of 8 sources because this is the average 
number of sources that are present in a grid cell.
The large distribution is extracted from an observed field centered on $l$=122\degr, 
$b$=1\degr. We have performed the same simulation using various other fields to test 
whether the results below change when less, or more reddening is present in a field.
We find that this is not the case.
 
\begin{figure*}[htbp]
\centering
\subfloat{
\includegraphics[width=1\linewidth,clip]{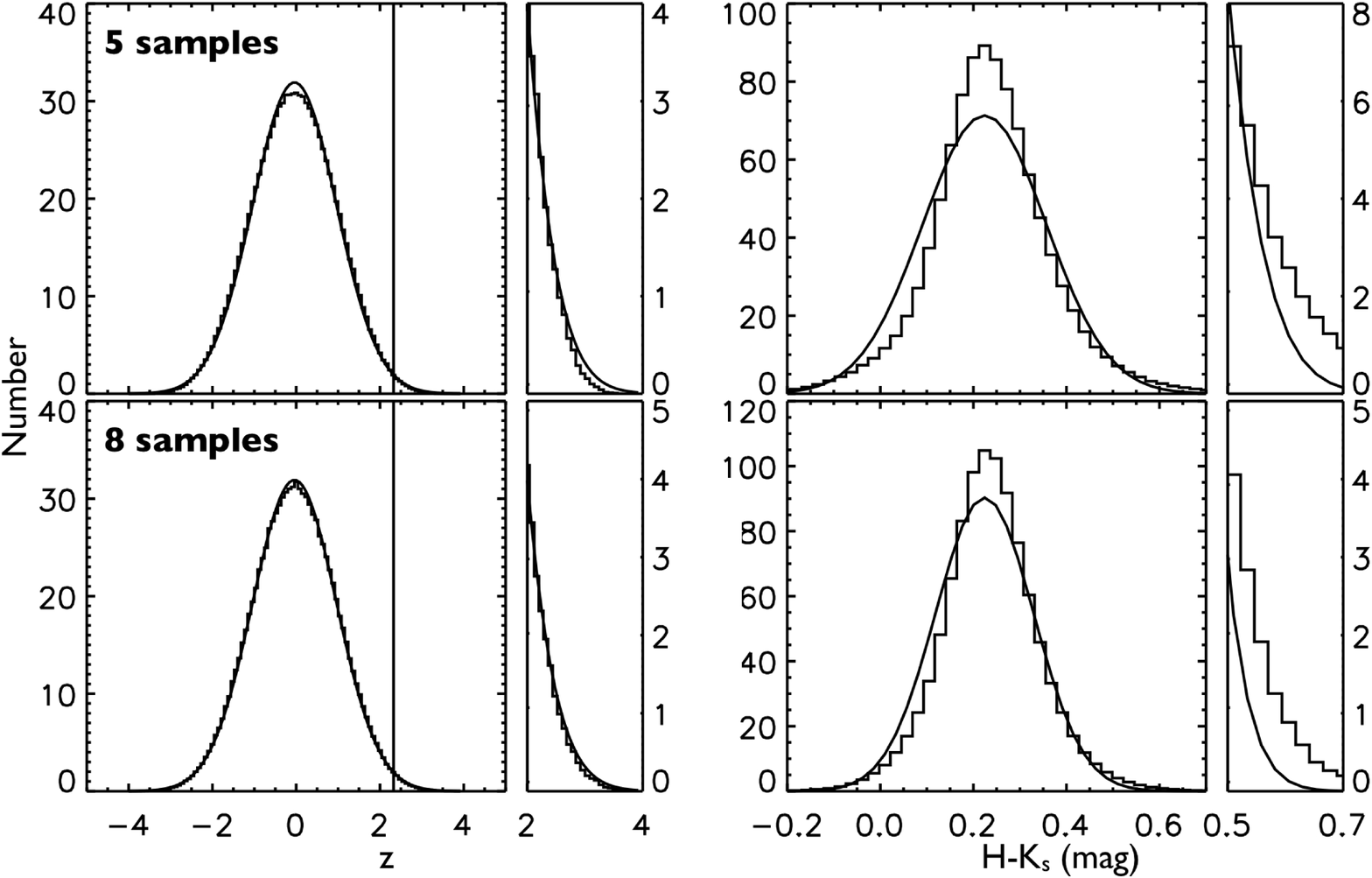}}
\caption{Visualization of the result of a Monte Carlo simulation, where we randomly 
draw 6000 samples of 5 (upper histograms) and 8 (lower histograms) 2MASS
colors from a test field at $l$=122\degr, $b$=1\degr. The distributions of the 
significance $z$ are displayed on the left. The right histogram displays the distribution 
of average colors in the same simulation. A blow-up of the high-end tail of each
distribution is given in the narrow panels right of the histograms.
The solid lines in each panel represent Gaussian profiles where the peak position and 
FWHM are adopted from the simulated distributions.}
  \label{fig_ave}
\end{figure*}

The resulting distribution of $z$ is given in the lower left panel in Figure \ref{fig_ave}.
In the upper left panel we display the distribution of $z$ for samples of 5 sources, the 
minimum sample size for a proper U-test result according to \citet{Moses1964}.
The peak as well as the wings are slightly overestimated by the Gaussian profile 
in the simulation for samples of 5 sources, but increasing to samples of 8 
shows that the simulation converges fast to a Normal distribution. 
The difference between the the areas under the curves for the wing region ($z>2$) is less 
than 10\% ($<$5\% for samples of 8).
By assuming that
the profile is given by a Gaussian distribution, we underestimate the actual probability
of having a red pixel in the regime considered here, i.e., $z\geq$\,2.326 
corresponding to a probability $P\geq$99\% when evaluating the integral over Equation 3.

We compared the simulated results of the U-test with two parametric tests: the average 
and median. The right panels of Figure \ref{fig_ave} show that the distribution of 
average values deviates much more from a Gaussian profile compared to the U-test.
The distribution is clearly skewed and cannot be represented by a simple profile. 
The area in the wing ($H-K_{\rm{S}}$$>0.5$\,mag) is underestimated by 
$\approx$50\% ($\approx$100\% for 8 samples). The shape of the distribution of average 
values approaches a Normal distribution only for sample sizes $\gtrsim20$, but the area 
in the wing remains underestimated.
Almost identical results are found for the distribution of median colors, with a 
comparably large discrepancy in the wing area.

The average and median colors are often used, e.g., for near-infrared color excess 
(NICE). Even though the median value is less sensitive to (extreme) outliers in the 
distribution than the average, the main issue is, that both are not well represented by
simple profiles. This makes it difficult, if not impossible, to put a significance to the 
measured values. This justifies the use of the non-parametric U-test statistics for the 
automated identification process presented in this paper.

\subsection{Reliability and Completeness}
There are several considerations that need to be kept in mind when interpreting the 
resulting images created by the U-test method. These are related to the 
reliability and completeness of the cells that are identified as being red. We discuss 
them here.\\
\ \\
{\it Type I error}\\
The first error is related to the reliability, or the significance level of the selected 
cells. If, as in our case, rejection is defined at 99\% this suggests that about 1\% of the 
cells will be selected on statistical arguments, rather than being red. However, these 
cells are expected to be randomly distributed over the fields. To exemplify that this is 
indeed the case, we consider the following values extracted from the low-resolution 
$2\degr \times 2\degr$ field toward $l$=166\degr, $b$=1\degr. This field is assumed to be 
empty, because no objects, 
identified as clusters of red cells, ended up in a final identification. However, the field 
contains 76 cells out of a total of 6561 ($\approx$ 1\%) where the Null hypothesis of the 
U-test was rejected. Indeed, these cells are spread more or less randomly over the field,
or else they would have been identified as clustered objects.\\
\ \\
{\it Type II error}\\
This error is related to the cells that are not identified, i.e., where the Null hypothesis 
is not rejected but should have been. It is very difficult to put a realistic value 
to this error. In a broad sense it is inversely related to the Type I error. That is, the 
more reliable the sample is the less complete it is and vice versa. Furthermore, the Type 
II error usually occurs when sample sizes are too small. Because we have chosen 
to produce a high reliability catalog, we are giving in on completeness due to this error.\\
\ \\
{\it Sample statistics}\\
According to Moses (1964), when one of the distributions that go into the U-test contains 
fewer than 5 sources the outcome of the U-test is unreliable. There is not sufficient 
information available on the color distribution. Cells containing fewer than 5 sources 
are omitted in the identification process and, thus, introduce another form of 
incompleteness which depends on the resolution. Processing at higher resolution
means being less complete.\\
\ \\
Considering the limitations and reliability of the method, we have decided to put
the effort in producing a reliable target list at relatively high resolution 
(60\arcs\ and 90\arcs), rather than being complete. 
Furthermore, our intention is not to evaluate the content or structure of the
outer galactic plane nor finding all 'dark clouds' present in the outer Galaxy. 
We purely present a statistical method that enables the identification of real objects.  
\section{Extracting candidate sources}
{\it Friends-of-friends approach}\\
To increase the reliability of red regions and to avoid the selection of random
cells, we add an additional step to the process. By means of a friends-of-friends
algorithm, we select only clusters of 4 or more red cells for the final list of candidate 
objects.
The reason for using 4 cells here is based on the probability of identifying
clusters of randomly distributed red cells, which becomes negligible ($<<$1\%) for groups
larger than 3 cells. These probabilities depend on the linking length described below
and were evaluated in the Monte Carlo simulations.
The friends-of-friends algorithm, first applied by \cite{Huchra1982} and often
encountered in cosmological studies, is a well-known technique and can be simplified to 
find groups, or structures, in projection on the sky. We use the friends-of-friends 
approach to identify groups of cells on the sky where the color distribution of stars is red
compared to the local surroundings, i.e., where the U-test returns a probability of
$P$= 99\% or higher.\\
\ \\
{\it Linking length}\\
A crucial parameter required for the friends-of-friend method is the linking length,
$L$, used to determine whether selected cells belong to the same group or not. If $L$
is taken too large, all cells will be assigned to the same group. If $L$ is too small all
cells will be considered isolated. Because we do not know the cause of the red
color distribution in
different cells (either extinction due to a foreground material at unknown distance, or a
concentration of intrinsically red objects at unknown distance) it is impossible without 
further information to put a physical size to $L$. We therefore choose a linking
length in terms of the cell size, $s_{\rm{cell}}$.

An upper limit for $s_{\rm{cell}}$ can be estimated from an empty field. Consistent with 
the defined reliability, about 1\% of the cells in an empty field are rejected by the 
U-test and spread randomly over the field. We assess an average distance between rejected 
cells by using a Monte Carlo simulation, where $N$ cells (=1\% of $a^2$) are selected at 
random locations from a square box with side $a$. For $a^2$ we use the number of cells that
make up a $2\degr \times 2\degr$ field at the given resolutions, i.e., $a$=81 cells at 
90\arcsec\ and $a$=121 cells at 60\arcsec\ resolution. The average distance in the 
simulation converges to 5.3 and 5.2 times the cell-size, respectively. This is in agreement 
with the theoretically expected value, given by
\begin{equation}
\rm{average\ distance} = \frac{1}{2}\sqrt{\frac{a\times a}{N}} = 5 \ \ \ [\times\  s_{\rm{cell}}] 
\end{equation}
\citep{Hertz1909,Clark1954}.

A linking length much larger than this average
distance will group cells together even if
the distribution is random. This obviously
contradicts the goal of finding clustered regions.
A lower limit is obviously the cell
size itself. We have tested the method by
varying $L$ between 2 and 5. The higher values
($>$3) identify mostly large scale structures
($>$0.5\degr). For a value of 2, most cells are
isolated or grouped with fewer than 3 other
cells. We decide to use a linking length half
that of the average distance between random
points, i.e., $L$=2.5 times the cell size, for
both resolutions. The probability for finding
'accidental groups' of 4 cells using $L$=2.5 and considering
that 1\% of the cells are randomly selected by the U-test
was evaluated in the Monte Carlo simulations 
described above and is $\lesssim$0.1\%. 

A consequence of setting the linking length to 2.5 is that groups are
identified with cells in between the red ones that may not be identified as red. This can either be due
to a Type II error (Sec. 3.6), or insufficient stars present in that cell (the sample 
statistics, Sec. 3.6). In both cases this illustrates that not all reddened regions
can be detected and thus implies that we are limited in the completeness of the catalog.
\section{Results}
\begin{figure*}[!ht]
\centering
\includegraphics[width=1\linewidth,clip]{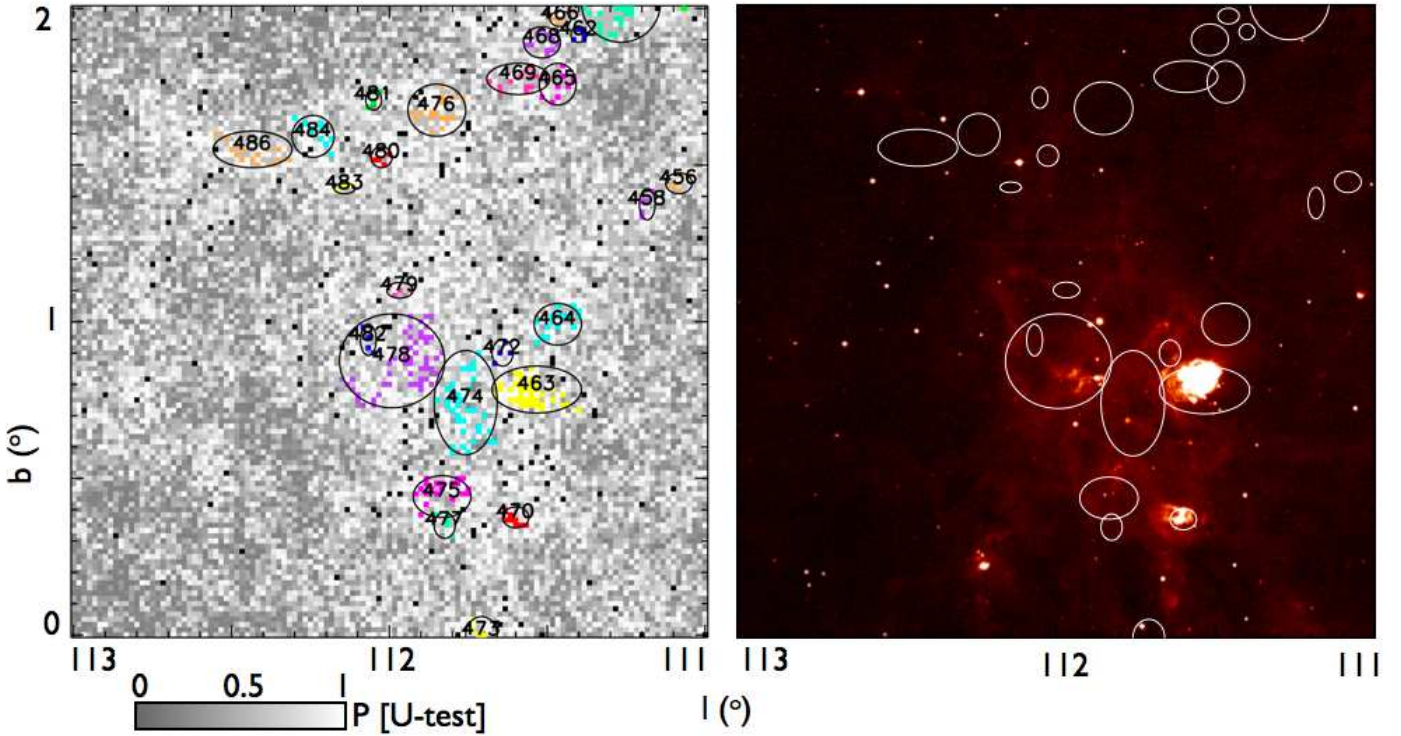}
\caption{The left panel displays the probability image returned by the U-test. The 
grey-scales represent the probability that the color distributions in the cells are 
redder with respect to the reference distribution. The friends-of-friends method 
identified the extended red regions, displayed in arbitrary colors. The ellipses 
give the spatial extent of the regions. Black cells are identified as being red, but 
are considered isolated. Objects 463 and 470 correspond to \object{NGC 7538} and \object{S 159}, 
respectively. Object 474 is the first outer Galaxy `cluster-forming clump'-candidate, 
identified as such by the work presented here. 
The right image shows the same region in MSX 8\,$\mu$m emission for comparison.
Bright emission from warm dust, heated by embedded, massive young stars, is seen toward 
both \object{NGC 7538} and \object{S 159}. The upper part of the image is almost devoid 
of extended 8\,$\mu$m emission, typical for the majority of directions in the outer Galaxy.}
  \label{fig_rel}
\end{figure*}

Figure \ref{fig_rel} displays an example of the output of the U-test in the form of a probability 
image for the region centered on $l$=112\degr, $b$=1\degr. For comparison we 
display the 8\,\mum\ emission as observed by MSX. We identify 24 extended red regions 
in this field. The red cells of these regions are presented in different colors 
in Fig. \ref{fig_rel}, with the spatial extent of the objects indicated by the ellipses.
Some objects clearly correlate with extended
bright 8\,\mum\ emission coming from well-known star forming regions such as \object{NGC 7538} or
\object{S 159}. This field also contains object H474, which has been selected as a `massive
dark cloud'-candidate for follow-up observations. These observations have revealed the 
first infrared dark cloud identified as such in the outer galactic plane 
\citep{Frieswijk2007,Frieswijk2008}.

The resulting catalog of extended red regions and the probability images in 
{\sc FITS}-format are available 
online\footnote{http://www.astro.rug.nl/$\sim$ismgroup/OuterGalaxy/}. The low-resolution
part consists of 1589 objects and the high resolution part of 1320.  The catalog is available
in electronic form at the CDS with sources designated FrSh LNNNN and FrSh HNNNN for low- 
and high resolution objects, respectively.
Table 1 presents an excerpt from the high resolution catalog. 
The objects in the table correspond to those that are identified in the field displayed in 
Figure 3 and include
\object{NGC 7538} (FrSh H463) and \object{S 159} (FrSh H470). The different 
columns present the following information:\\
\ \\
Column (1): Object identification\\
Column (2)--(3): Galactic coordinates of the object centre\\
Column (4): Total number of red pixels\\
Column (5)--(6): Extent in $l$ and $b$ in pixel units (1 pix=1\arcm)\\
Column (7): Number of 2MASS sources\\
\ \\
Note that the number of 2MASS sources in Column (7) contains information
on the stellar surface density. In principle this can be used to distinguish between
objects identified by extinction and objects representing a clustering of 
young stars, because extinction may lower the stellar density whereas a
clustering may increase it. However, because we are using the 2MASS catalog
including the faint extension the variations in stellar surface density may
also arise from the incompleteness of the catalog and we do not discuss it 
further in this paper.

\renewcommand{\footnoterule}{}  
\begin{table}[!ht]
	\begin{minipage}[t]{\columnwidth}
	\small
	\label{tab:3_tab_cat}
	\centering
	\renewcommand{\footnoterule}{}  
	\begin{tabular}{ccccccc}
\hline\hline 
  \multicolumn{1}{c}{FrSh} &
  \multicolumn{1}{c}{l} &
  \multicolumn{1}{c}{b} & 
  \multicolumn{1}{c}{N$_{cell}$}&
  \multicolumn{1}{c}{$\Delta l$}&
  \multicolumn{1}{c}{$\Delta b$}&
  \multicolumn{1}{c}{2M}\\
  
\multicolumn{1}{c}{(1)}& 
\multicolumn{1}{c}{(2)}& 
\multicolumn{1}{c}{(3)}&
\multicolumn{1}{c}{(4)}& 
\multicolumn{1}{c}{(5)}& 
\multicolumn{1}{c}{(6)}& 
\multicolumn{1}{c}{(7)}\\
  \multicolumn{1}{c}{ }& 
  \multicolumn{1}{c}{deg}&
  \multicolumn{1}{c}{deg}& 
  \multicolumn{1}{c}{ }& 
  \multicolumn{1}{c}{pix}& 
  \multicolumn{1}{c}{pix}& 
  \multicolumn{1}{c}{}\\
\hline
H456  &    111.08  &      1.44  &        5  &        5  &        4  &       43 \\ 
H457  &    111.27  &      2.02  &       41  &       15  &       15  &      299 \\ 
H458  &    111.18  &      1.38  &        5  &        3  &        6  &       49 \\ 
H459  &    111.20  &      2.17  &        6  &        3  &        5  &       51 \\ 
H460  &    111.22  &     -0.83  &        5  &        3  &        5  &       79 \\ 
H461  &    111.27  &     -0.75  &        4  &        3  &        5  &       68 \\ 
H462  &    111.40  &      1.92  &        5  &        3  &        3  &       38 \\ 
H463  &    111.53  &      0.78  &       40  &       17  &        9  &      655 \\ 
H464  &    111.47  &      0.99  &       16  &        9  &        8  &      122 \\ 
H465  &    111.47  &      1.76  &        9  &        7  &        8  &       67 \\ 
H466  &    111.46  &      1.97  &        4  &        4  &        3  &       29 \\ 
H467  &    111.48  &     -0.66  &        5  &        5  &        2  &       78 \\ 
H468  &    111.52  &      1.89  &       10  &        7  &        6  &       70 \\ 
H469  &    111.59  &      1.77  &       13  &       12  &        6  &       80 \\ 
H470  &    111.60  &      0.38  &        8  &        5  &        4  &      123 \\ 
H471  &    111.62  &      2.17  &        4  &        6  &        2  &       24 \\ 
H472  &    111.64  &      0.90  &        4  &        4  &        5  &       32 \\ 
H473  &    111.71  &      0.00  &        8  &        6  &        7  &       84 \\ 
H474  &    111.76  &      0.74  &       34  &       12  &       20  &      335 \\ 
H475  &    111.83  &      0.44  &       20  &       11  &        8  &      162 \\ 
H476  &    111.85  &      1.67  &       16  &       11  &       10  &      122 \\ 
H477  &    111.82  &      0.35  &        5  &        4  &        5  &       46 \\ 
H478  &    111.99  &      0.88  &       52  &       20  &       18  &      468 \\ 
H479  &    111.97  &      1.10  &        4  &        5  &        3  &       31 \\ 
H480  &    112.03  &      1.52  &        5  &        4  &        4  &       53 \\ 
H481  &    112.05  &      1.71  &        4  &        3  &        4  &       29 \\ 
H482  &    112.07  &      0.94  &        4  &        3  &        6  &       34 \\ 
H483  &    112.14  &      1.42  &        4  &        4  &        2  &       41 \\ 
H484  &    112.24  &      1.59  &       11  &        8  &        8  &      101 \\ 
H485  &    112.32  &      2.84  &        7  &        5  &        6  &       51 \\ 
H486  &    112.43  &      1.55  &       15  &       15  &        7  &      109 \\ 
\hline
\end{tabular}
\caption{Excerpt from the catalog of extended red objects.}
\end{minipage}
\end{table}

\subsection{Size- and spatial distribution}
\label{spatial}
\begin{figure}[ht!]
  \centering
\includegraphics[width=0.5\textwidth,angle=0]{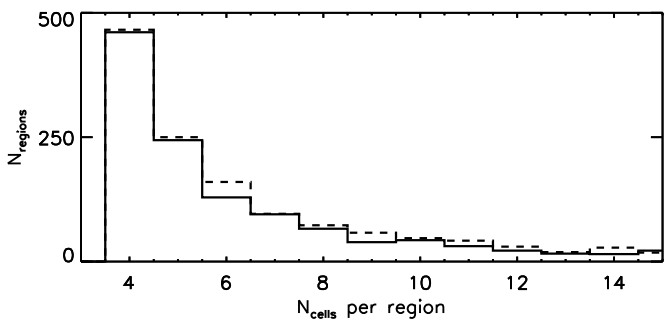}
\caption{Histogram displaying the distribution of the number of cells per extended 
red region for the high-resolution (solid) and low-resolution (dashed) catalog.}
  \label{fig_size}
\end{figure}

The linear extent of the extended red regions ranges from a few arc-minutes to
about a degree. Figure \ref{fig_size} gives an overview of the cell-number
distribution for the regions in the high- and low resolution catalog. 
This shows that about half of the regions are identified as groups of 4 or 5 cells. 
Regions consisting of more than 10 cells account for 18\% of the objects in the high
resolution catalog and for 28\% of the objects in the low resolution one.

\begin{figure*}[ht!]
  \centering
\includegraphics[width=1\textwidth,angle=0]{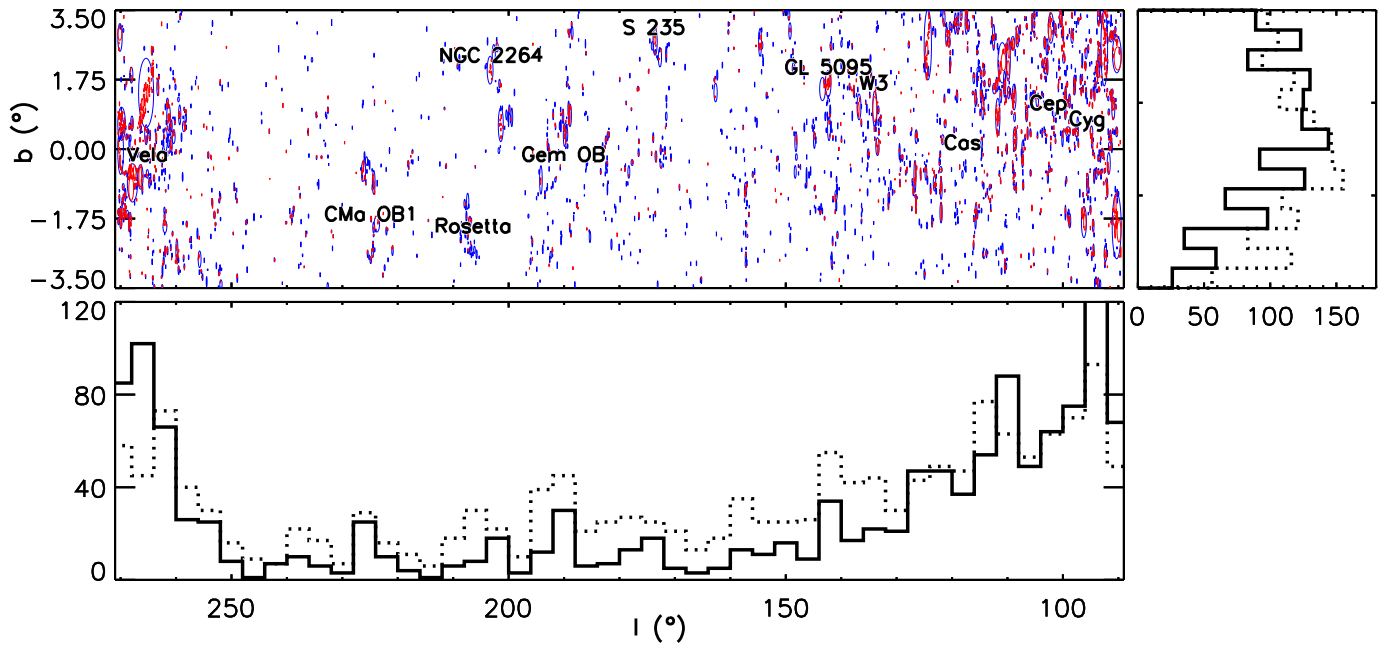}
\caption{Spatial distribution of extended red regions identified in the outer galactic 
  plane at 60\arcsec\ (solid histogram, red regions) and 90\arcsec\ (dashed histogram, 
  blue regions) resolution. The regions are identified with a reliability of $>$99.9\%.
  A few well-known star forming regions in the Galaxy are given for reference.}
  \label{fig_spatial}
\end{figure*}
Figure \ref{fig_spatial} shows the sky distribution of the extended red regions in the 
outer galactic plane.
The solid histogram in the lower panel represents the number of high-resolution regions 
in bins of 5\degr\ galactic longitude. The dashed histogram displays
the distribution of low-resolution regions. 
The galactic latitude distribution is given in the horizontal histogram  
in the right panel. Several authors have found that the distribution of (molecular) clouds
toward the inner galactic plane peaks at negative galactic latitude 
\citep{Peretto2009,Schuller2009,Rosolowsky2009}, suggested to result from the Sun being 
located slightly above the Galactic plane.
We do not find a similar distribution toward the outer Galaxy. In fact, about 60\% of the
high resolution objects are at positive latitudes ($\approx$50\% for low resolution). 
However, this can be attributed to the local, well-populated, Vela and Cygnus/Cepheus star 
forming regions which are located mainly above the plane.

The 2-dimensional spatial distribution is 
displayed in the upper panel. The red and blue ellipses display the location and linear 
extent of the high and low resolution regions, respectively.
Some well-known areas associated with molecular clouds and star forming activity in the 
outer Galaxy are indicated. Most of the regions we find are well confined within the extent 
of these large complexes, and by far most regions are identified in
the crowded spiral arm regions near Vela and Cygnus/Cepheus.
Toward lower stellar densities ($l\approx 140\degr$--$240\degr$) there are more 
regions identified with the low resolution than the high resolution grid because at
high resolution there are often insufficient stars in a cell to use for the U-test.
Many of the large-scale regions identified at low resolution, in particular visible toward 
the Vela and Cygnus/Cepheus regions, are identified as groups of smaller regions with the 
high resolution grid.

\subsection{Column density sensitivity}
\label{sensitivity}
The objects in the catalog are identified without using a parameter that relates to a
physical quantity. However, we do have information on the average \hk\, colors of cells in 
all objects and we use it to derive a rough estimate of the column density sensitivity.
Note however, that we can not account for any corrections due to contamination of unreddened
foreground stars and the values below represent only lower limits.

The average color in the cells of identified regions is $<$\hk$>$$\approx$0.54 mag 
with a standard deviation of $\sigma$$\approx$0.18 mag. Some cells have an average color 
up to $\approx$2 mag.
These value can be converted to a visual extinction $A_{\rm V}$ using 
the following equations;
\begin{equation}
	A_{\rm{V}} = 15.9 \times E(H-K_{\rm{S}}) \ \ \ 
\end{equation}
\citep{Rieke1985},
\begin{equation}
	E(H-K_{\rm{S}}) = (H-K_{\rm{S-observed}}) - (H-K_{\rm{S-intrinsic}})\ ,
\end{equation}
where $E(H-K_{\rm{S}})$ represents the color excess, $(H-K_{\rm{S-observed}})$ the
observed color in a cell and $(H-K_{\rm{S-intrinsic}})$ the intrinsic color. For the 
intrinsic color we use a value of 0.15\,mag (Sec. \ref{data}).
This means we are sensitive to visual extinction values ranging from a few to $\approx$30 
magnitudes, corresponding to molecular column densities up to
N(H$_2$)$\approx$3$\times$10$^{22}$\,cm$^{-2}$ \citep[][]{Bohlin1978}. 
This is comparable to values found for the majority of IRDCs \citep{Peretto2009}.
\section{Conclusions}
\label{conclusions}
We have used a statistical approach in combination with a friends-of-friends algorithm
to measure deviations in the spatial \hk\ color distribution of stars in the outer 
galactic plane, with the goal to identify extended reddened regions.
We processed the galactic plane at 60\arcsec\ and 90\arcsec\ resolution, resulting
in the identification of 1320 high resolution and 1589 low resolution extended red 
regions. The reliability of individual red cells that make up each object is 99\% and 
by using a friend-of-friends approach, the resulting catalog of objects is 99.9\% 
reliable. Because we have put the effort in producing a highly reliable source list
we cannot quantify the completeness of the catalog.

The majority of the objects are located toward well-known molecular cloud complexes and 
some correspond to specific, well-known objects such as \object{NGC 7538} and \object{S 159}. This 
correlation strengthens the argument that the red nature of the regions is caused by 
extinction and/or embedded young stellar objects rather than intrinsic star color variations.
The main goal of our analysis is to find previously undetected dark clouds
in the outer Galaxy. However, many other types of objects are included in the catalog 
and the results may be compared to other outer Galaxy studies in the future.

Based on the parameters derived from the statistical test it is impossible to
determine the nature of the reddening. The next step is to cross-correlate the
objects with existing optical, infrared and CO data. This process is described
in an accompanying paper (Frieswijk et al. 2010).
\begin{acknowledgements}
	We thank the anonymous referee for his/her care- 
	ful reading of the manuscript and his/her constructive remarks. 
	We would also like to thank Marco Spaans and Floris van der Tak for their 
	helpful discussions and suggestions which have improved this manuscript.
	This publication makes use of data products from the
	Two Micron All Sky Survey, which is a joint project of the
	University of Massachusetts and the Infrared Processing and Analysis
	Center/California Institute of Technology, funded by the National 
	Aeronautics and Space Administration and the National Science Foundation.
\end{acknowledgements}
\bibliographystyle{aa}
\bibliography{master}
\end{document}